          [1999/12/02 1.01 (PWD)]
\documentclass[a4paper,twocolumn]{esapub} % European paper

\usepackage{epsfig}
\usepackage{natbib}
\title{The COMPTEL 1.809 MeV Survey}

\author[1]{S. Pl\"uschke}
\author[1]{R. Diehl}
\author[1]{V. Sch\"onfelder}
\author[2]{H. Bloemen}
\author[2]{W. Hermsen}
\author[3]{K. Bennett}
\author[3]{\\C. Winkler}
\author[4]{M. McConnell}
\author[4]{J. Ryan}
\author[5]{U.Oberlack}
\author[6]{J.Kn\"odlseder}
\affil[1]{Max-Planck-Institut f\"ur extraterrestrische Physik, 85740 Garching,
Germany}
\affil[2]{SRON, 3584 CA Utrecht, The Netherlands}
\affil[3]{Astrophysics Division, ESTEC, 2200 AG Noordwijk, The Netherlands}
\affil[4]{Space Science Center, Univ. of New Hampshire, Durham, NH 03824, USA}
\affil[5]{Astrophysics Laboratories, Columbia University, New York, NY 10027, USA}
\affil[6]{CESR, CNRS/UPS, 31028 Toulouse, France}

\newcommand{\al}{\ensuremath{^{26}{\rm Al}} }

\begin{document}


\maketitle

\begin{abstract}
We present the latest update of the 1.809 MeV sky survey obtained with COMPTEL.
Based on all observations taken since the launch of CGRO in spring 1991 to early
summer this year we obtain 1.809 MeV all sky maps using different imaging methods.
The background is modelled on the basis of an adjacent energy approach.
We confirm the previously reported characteristics of the galactic 1.809 MeV
emission, specifically excesses in regions away from the inner Galaxy.\\
The observed 1.8 MeV $\gamma$-ray line is ascribed to the radioactive decay of
$^{26}$Al in the interstellar medium. $^{26}$Al has been found to be
predominantly synthesised in massive stars and their subsequent core-collapse
supernovae, which is confirmed in tracer comparisons. Due to this, one
anticipates flux enhancements aligned with regions of recent star formation,
such as apparently observed in the Cygnus and Vela regions.
\end{abstract}
\section{Introduction}
The imaging gamma-ray telescope COMPTEL \citep{sch93} aboard NASA' s CGRO spacecraft
allowed for the first time to survey the entire sky in the MeV regime. One of the
mission highlights is the generation of the first all-sky images of the 1.809 MeV
gamma-ray line emission, first detected by \cite{mahoney82} using HEAO-C. This emission
line is attributed to the radioactive decay of \al with a lifetime of 1.04 Myr.
The first 1.809 MeV all-sky map based on the first three years of COMPTEL
observations was presented by \cite{oberlack96}. This image was based on a
maximum entropy deconvolution method (ME) \citep{strong92} applying an adjacent
energy background model \citep{oberlack96,knoedl96} to the individual observation
periods. This map confirmed the non-local character of the detected 1.809 MeV emission,
as already seen in a first Galactic plane survey \citep{diehl95} and confirmed later
in the 5 year image of \cite{oberlack97}. We attribute most of the emission to young,
massive stars and active star forming regions.\\
The maximum entropy deconvolution shows a tendency for highly structured images
\citep{oberlack97, knoedl97}. Other image reconstruction methods have been developed
and applied to the COMPTEL data. \cite{knoedl99} introduced a multi-resolution
regularised expectation maximisation algorithm (MREM) using a wavelet filtering
for noise suppression. Although the MREM images are much less structured than the
ME maps, the previously reported emission characteristics are confirmed. Possibly
the MREM approach is somewhat conservative with respect to image structures whereas
the maximum entropy images still may include artifacts.\\
In addition to these imaging approaches, which make use of the adjacent energy
background described below, \cite{bloemen99} showed an alternative approach. 
They use an iterative model fitting method combined with a maximum entropy imaging
of the residual emission in neighbouring energy bands to construct an appropriate
background model, which is finally used for the maximum entropy imaging in the line
energy regime. A comparison to the cycle 1-5 image \citep{oberlack97}, from the
same COMPTEL data, shows an underestimation of the deduced fluxes by applying the
adjacent energy background instead of the iterative modell fitting. Yet, the
reported image structures are rather identical.\\
Application of maximum entropy and MREM imaging to seven years of COMPTEL data
with a refined adjacent energy background model again confirms the chief image
features \citep{plueschke00}. After termination of the CGRO mission end of May 
this year we now present results using the complete mission database. The following
section summarises the 1.809 MeV data processing, followed by a presentation
of the maximum entropy and multi-resolution images. Finally we discuss tests of
possible systematic effects on the imaging results.
\section{Data Analysis \& Background Treatment}
In this analysis we use all data from beginning of the mission up to its end,
from May 1991 to end of May 2000, split into $\sim 350$ observation periods with
typical durations of one to four weeks. The accumulated effective observation time
ranges from 0.27 to $4.20\cdot 10^{7}\,{\rm s}$, which results in a sensitivity in
the 1.809 MeV regime of 0.8 to $1.4\cdot 10^{-5}\,{\rm ph\,cm^{-2}\,s^{-1}}$,
depending on location.\\
For the imaging analysis the event data from a 200 keV wide energy band around
the 1.8 MeV line are binned in a 3-dimensional data space, which is spanned by
scatter direction of the photon between the two detector layers inside the telescope
and the scatter angle resulting from the Compton kinematics. A more detailed
description of the data space and the event selection criteria can be found in
\cite{oberlack97}. The instrument characteristics in the energy regime are 
summarised in a energy resolution of 140 keV (FWHM) and an angular resolution
of $3.8^{\circ}$ (FWHM).\\
Instrumental background strongly disturbs the observation data. The effective
signal to noise ratio is of the order 1\%, so accurate background modelling is 
crucial in analysing COMPTEL data. In the case of gamma-ray line analysis the
use of adjacent energy bands for background deduction appears to be natural. To
account for time dependent variations of the background the model is estimated 
on an observation period basis and summed afterwards. To suppress statistical
fluctuations each background model is smoothed by a local chi-square fit to the
geometry function. This accounts for the original event selection characteristics
of the specific observation periods and guarantees an accurate treatment also in
the galactic polar region.\\
A longterm study by \cite{oberlack97} of count-ratios in the line energy band relative
narrow adjacent energy bands showed a clear time-dependence. This time-variability
can be traced to a build up $^{22}$Na during the mission, affecting the spectral shape
in the vicinity of the $^{26}$Al line.
\cite{weiden00} identified instrumental background components due to 8 radioactive isotopes,
with $^{22}{\rm Na}$ and $^{24}{\rm Na}$ being the strongest among them. 
%A detailed study of COMPTEL´s instrumental background by \cite{weiden00} demonstrated the
%origin $^{22}{\rm Na}$ and $^{24}{\rm Na}$ variations, and their cascade lines to fall in
%the used energy regime. 
Both isotopes originate from activation in the structure elements of COMPTEL and CGRO.
The longer lived isotope $^{22}{\rm Na}$ shows a build up whereas $^{24}{\rm Na}$ is
responsible for short-time variations. \cite{oberlack97} determined the contributions of
these two isotopes per observation period by fitting the spectra using appropriate
templates. During the early mission this procedure restores the background normalisation
quite well.
%Beside these two isotopes \cite{weiden00} report on the identification of six additional
%isotopes contributing to the instrumental background. Especially in 
In the later mission phases, especially after the second reboost of CGRO spacecraft, the
additional isotopes must be taken into account to obtain an adequate representation. This
is because of the more efficient activation in the higher altitude orbits.
\begin{figure}[h!]
  \begin{center}
    \epsfig{figure=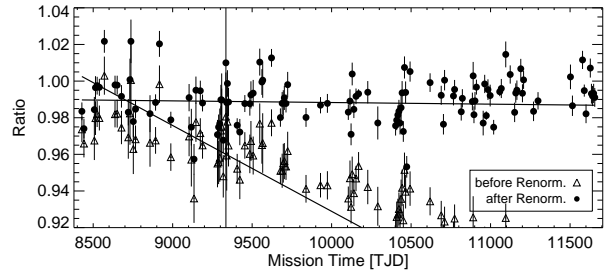,width=8cm}
    \caption{Time evolution of the background normalisation with (dots) and
             without (triangles) instrumental background correction.}
    \label{fig:bgd}
  \end{center}
\end{figure}
Figure \ref{fig:bgd} shows the background normalisation as function of the mission
time for both cases.
\section{The 1.809 MeV Maps}
\begin{figure}[h!]
  \begin{center}
    \epsfig{figure=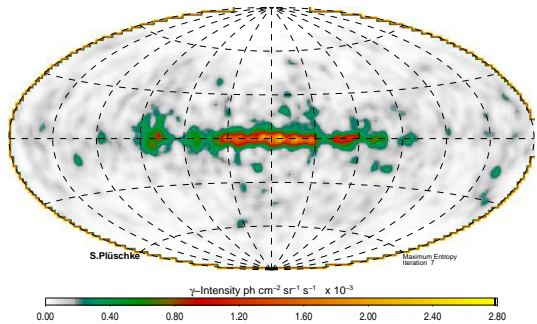,width=8cm}
    \caption{Maximum entropy all-sky image of the galactic 1.809 MeV emission
             observed with COMPTEL over 9 years.}
    \label{fig:me}
  \end{center}
\end{figure}

\begin{figure}[h!]
  \begin{center}
    \epsfig{figure=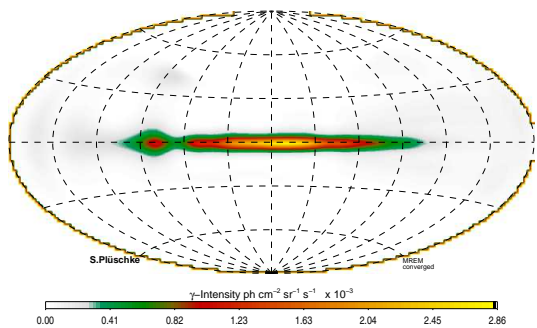,width=8cm}
    \caption{MREM all-sky image of the galactic 1.809 MeV emission
             observed with COMPTEL over 9 years.}
    \label{fig:mrem}
  \end{center}
\end{figure}

Following the analysis of \cite{plueschke00} we applied both image reconstruction
methods to our prepared data. We used the maximum entropy algorithm (see
\cite{strong92}), which iteratively extracts the sky intensity distribution
being compatible with the data. As already reported earlier, the ME method shows
a clear tendency to create a lumpy, structured image in late iterations. On the
other hand the early iterations significantly underestimate the gamma-ray fluxes.
Therefore an intermediate iteration has been chosen as a compromise between
flux reproduction and map smoothness (see figure \ref{fig:me}).\\
Alternatively, we applied the MREM technique to our data. The MREM algorithm
is based on an iterative expectation maximisation scheme accompanied by a
wavelet filtering algorithm. This wavelet filter suppresses the features of
low significance and artifacts by applying a user-adjustable threshold. By
controlling the changes in the reconstructed flux distribution this method becomes
convergent. The MREM algorithm attempts to produce the smoothest image being
consistent with the given data (see \cite{knoedl97,knoedl99}). Figure \ref{fig:mrem}
shows the equivalent MREM image of the COMPTEL data.\\
Both image reconstructions show an extended galactic ridge emission 
mostly concentrated towards the galactic center region ($|l|\le30^{\circ}$),
plus an emission feature in the Cygnus region, and a low-intensity
ridge along Carina and Vela. These characteristics confirm again the
previously reported emission structures. In addition the maximum entropy
image shows some low-intensity features in the longitude range between
$110^{\circ}$ and $270^{\circ}$. Also at latitudes beyond $\pm 30^{\circ}$
some of these features are visible. These features and their significance
are subject to further studies.
\section{Systematics}
\begin{figure}[t!]
  \begin{center}
    \epsfig{figure=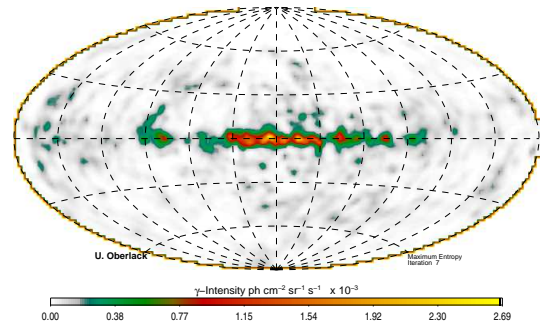,width=8cm}
    \caption{Maximum entropy all-sky image of the galactic 1.809 MeV emission
             from \cite{oberlack97}}
    \label{fig:ugo}
  \end{center}
\end{figure}

Figure \ref{fig:ugo} shows the maximum entropy image generate by \cite{oberlack97}
from the first five years of COMPTEL observations.
A comparison with figure \ref{fig:me} reveals only minor differences. The two
most obvious differences are the variation of the shape of the Cygnus feature
and the broader appearance of the emission in the complete mission image.\\
Due to the large field of view the COMPTEL observations covered the full
sky and accumulated very long effective observation times for each sky pixel.
Nevertheless the effective observation time varies over one order of magnitude.
Even when the point spread function is taken into account the exposure varies
within a factor of 2. These imperfections of the exposure may affect the
imaging results due to existing gradients \citep{oberlack97}. To investigate
these possible effects we selected observation periods so that the summed data
gives an exposure as even as possible. Due to a rather uneven
\begin{figure}[h!]
  \begin{center}
    \epsfig{figure=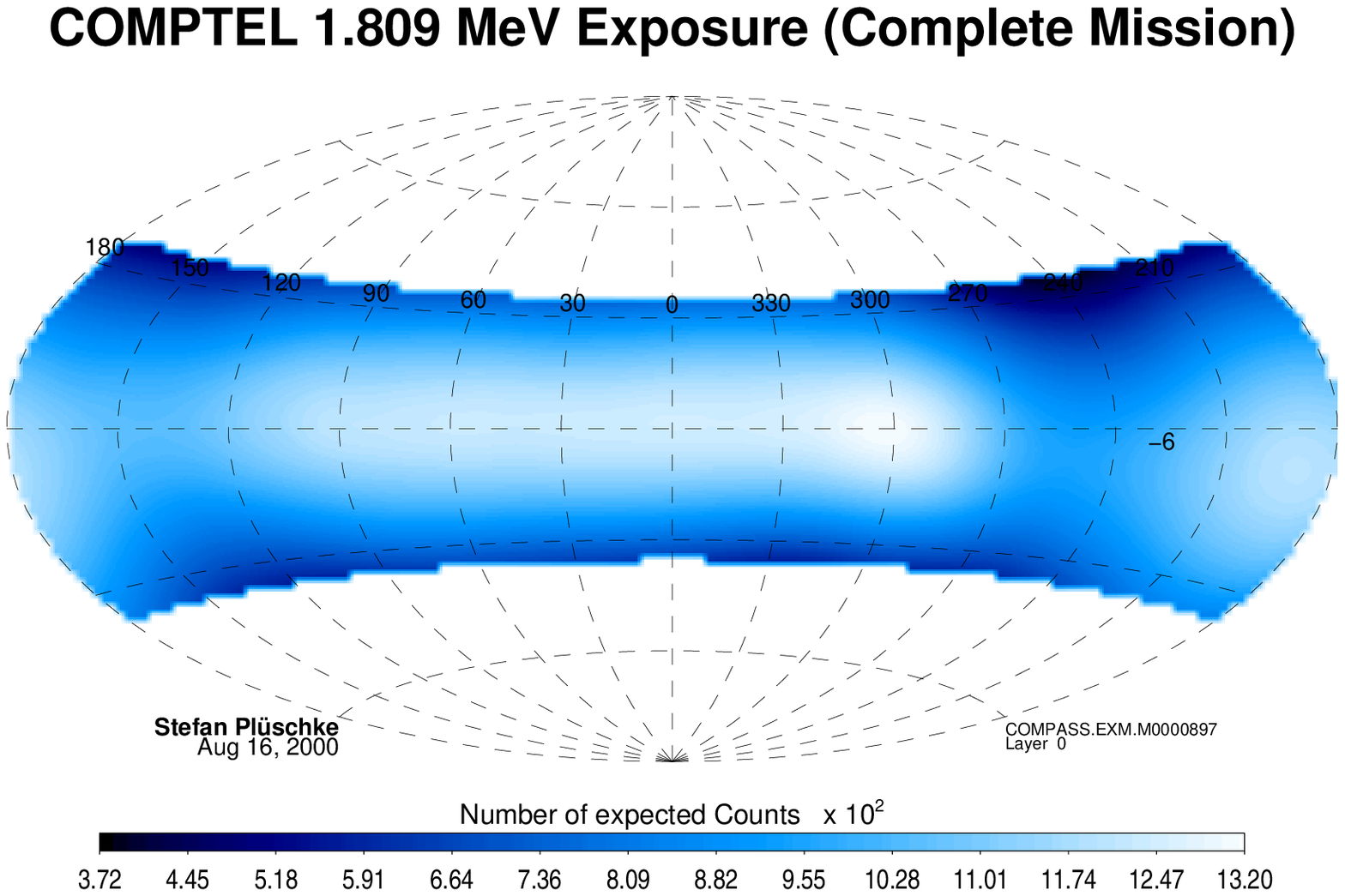,clip=,width=8cm}
    \epsfig{figure=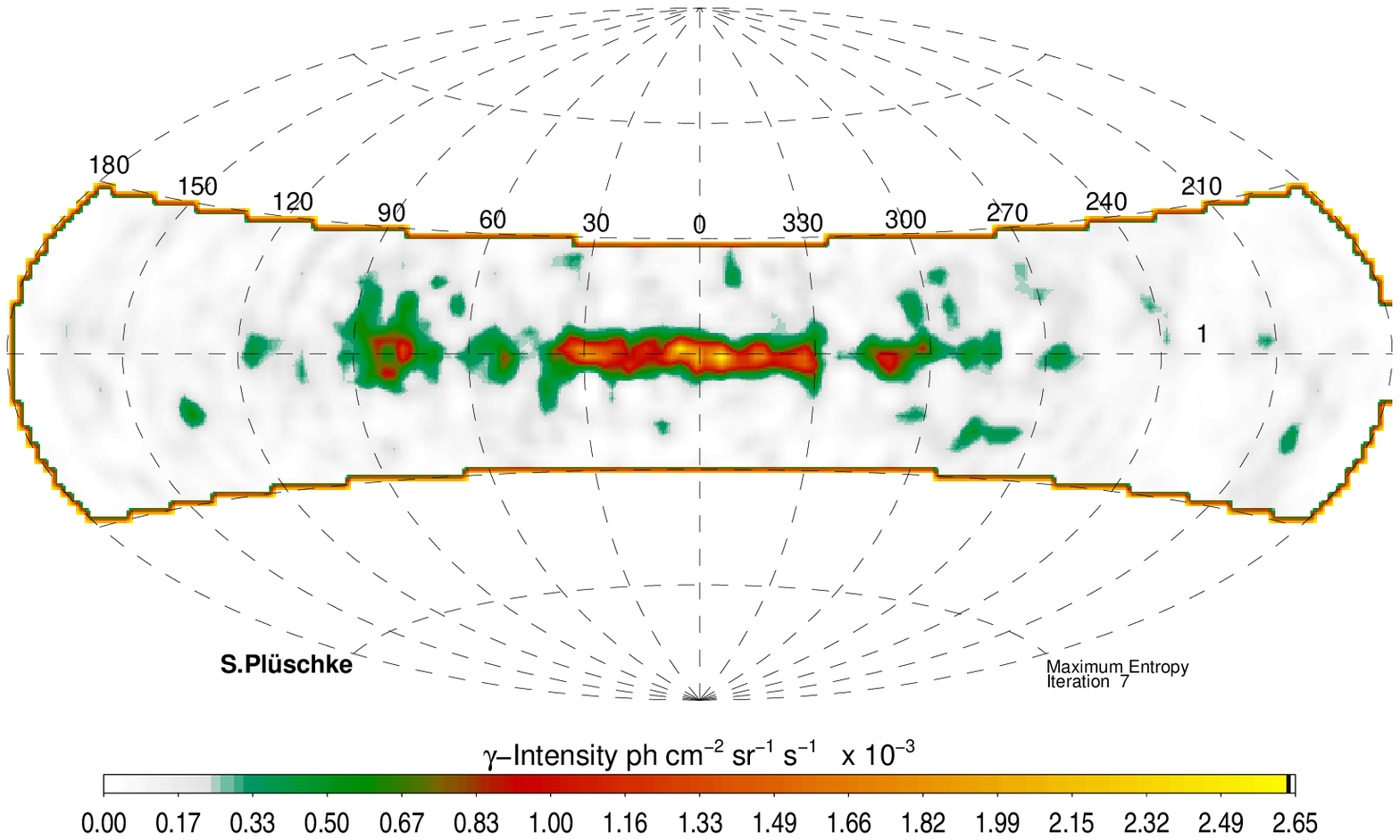,clip=,width=8cm}
    \caption{The upper panel shows the exposure resulting from selected observation
             periods. The goal is to constuct a data sample of an exposure being as even
             and homogeneuos as possible. The lower panel displays the resulting maximum
             entropy image of the galactic plane.}
    \label{fig:exp}
  \end{center}
\end{figure}
exposure near the northern galactic pole compared to the southern
hemisphere this is only possible for a band along the galactic equator
with $|b|\le40^{\circ}$. A further restriction is given by the fact that
at two positions in the galactic plane near $l\approx140^{\circ}$
and $l\approx250^{\circ}$ the exposure is only filled due to the large
field of view, no real observations pointing in these directions have
been undertaken. So a small gradient in the selected data still remains.\\
The upper panel of figure \ref{fig:exp} shows the exposure resulting
from the selected observation periods whereas the lower panel shows the
deduced maximum entropy image of these observations. A comparison of
the resulting image with figure \ref{fig:me} reveals no real differences
in the appearing features. Only the latitude extent of the selected data
image appears smaller than in the complete mission map, which could
be understood in perspective of the limitation of the selected viewing
periods to be restricted to pointings near the galactic plane.
%In this way the map is more similar to the 5-year image of \cite{oberlack97}
%given in figure \ref{fig:ugo}. Especially in the late mission phase
%COMPTEL was more often pointed towards the high latitude regimes,
%which may explain the latitude extension difference.
\section{Summary}
We have presented the COMPTEL 1.809 MeV all-sky maps based on the complete
mission database. COMPTEL reached an accumulated sensitivity in the 1.809 MeV
regime of 0.8 to $1.4\cdot10^{-5}\,{\rm ph\,cm^{-2}\,s^{-1}}$. The maps confirm
the previously reported emission characteristics - an extended ridge concentrated
towards the galactic center, a peculiar emission feature in the Cygnus region
\citep{plueschke00b} and a low-intensity ridge in the Carina-Vela \citep{diehl00}
region. 
In addition, the maximum entropy deconvolution gives some low-intensity features
which are suppressed in the MREM images. These features may be artifacts. A small
number of these features on the other hand appear in all earlier maximum entropy
maps at the same position, which gives some confidence in the reality of these
emission features. INTEGRAL will possibly allow a further study of these features.
\section*{Acknowledgements}
The COMPTEL project is supported by the German 'Ministerium 
fuer Bildung und Forschung' through DLR grant 50 QV 90968.

\end{document}